\documentclass[twocolumn,superscriptaddress,floatfix,preprintnumbers]{revtex4}
\usepackage{amsmath, amsfonts}    % need for subequations
\usepackage{graphicx}   % need for figures
\usepackage{verbatim}   % useful for program listings
\usepackage{color}      % use if color is used in text
\usepackage{subfigure}  % use for side-by-side figures
\usepackage{hyperref} %ext links, includi  % use for hypertng those to external documents and URLs
\usepackage{natbib}
\usepackage{siunitx}
\usepackage{physics}
\usepackage{ulem}
\usepackage[percent]{overpic}
\usepackage{booktabs}

\usepackage{xspace}
\usepackage[dvipsnames]{xcolor}

\usepackage[acronym]{glossaries}
\makeglossaries

\newcommand{\ignore}[1]{ }

\newacronym{AMO}{AMO}{atomic, molecular, and optical physics}

\newacronym{CPMG}{CPMG}{Carr-Purcell-Meiboom-Gill} %new

\newacronym{ESR}{ESR}{electron spin resonance}

\newacronym{FID}{FID}{free induced decay}

\newacronym{NV}{NV}{nitrogen-vacancy}

\newacronym{ODMR}{ODMR}{optical detected magnetic resonance} %new

\begin{document}

\title{
Search of High-Frequency Variations of Fundamental Constants \\ Using  Spin-based Quantum Sensors
}

\author{Xi Kong}
\altaffiliation{These authors contributed equally to this work.}
\author{Yuke Zhang}
\altaffiliation{These authors contributed equally to this work.}
\author{Chenyu Ji}
\altaffiliation{These authors contributed equally to this work.}
\author{Shuangju Chang}
\altaffiliation{These authors contributed equally to this work.}
\affiliation{The State Key Laboratory of Solid State Microstructures and Department of Physics, Nanjing University, 210093 Nanjing, China}

\author{Yifan Chen}
\affiliation{Niels Bohr International Academy, Niels Bohr Institute, Blegdamsvej 17, 2100 Copenhagen, Denmark}

\author{Xiang Bian}
\affiliation{The State Key Laboratory of Solid State Microstructures and Department of Physics, Nanjing University, 210093 Nanjing, China}

\author{Chang-Kui Duan}
\affiliation{CAS Key Laboratory of Microscale Magnetic Resonance and School of Physical Sciences, University of Science and Technology of China, Hefei 230026, China}
\affiliation{Hefei National Laboratory, University of Science and Technology of China, Hefei 230088, China}

\author{Pu Huang}
\email{hp@nju.edu.cn}
\affiliation{The State Key Laboratory of Solid State Microstructures and Department of Physics, Nanjing University, 210093 Nanjing, China}

\author{Jiangfeng Du}
\email{djf@ustc.edu.cn}
\affiliation{CAS Key Laboratory of Microscale Magnetic Resonance and School of Physical Sciences, University of Science and Technology of China, Hefei 230026, China}
\affiliation{Hefei National Laboratory, University of Science and Technology of China, Hefei 230088, China}
\affiliation{Institute of Quantum Sensing and School of Physics, Zhejiang University, Hangzhou 310027, China}

\maketitle

Precision searches of fundamental constant variations significantly enhance our understanding of the natural world. Earlier research predominantly utilized atomic spectroscopy \cite{Zhang:2022ewz} and optomechanical systems \cite{Vermeulen:2021epa} to search variations below 0.1 GHz. However, investigating high-frequency variations in fundamental constants remains an experimental challenge, despite its significance for many fundamental physics questions. We propose and implement an experiment harnessing  spin quantum sensor to search high-frequency variations in fundamental constants, encompassing the fine structure constant and electron mass, in the previously uncharted frequency domain from 0.1 to 12 GHz. This approach yields constraints on their relative variations as low as 5 parts per million (ppm)  and 8 ppm, respectively. Furthermore, based on these results, we establish stringent upper limits on the coupling constants associated with scalar field dark matter from 0.4 $\mu$eV to 50 $\mu$eV. Our research highlights the potential of  spin quantum sensors to investigate fundamental physics at high frequencies. Although our sensitivity requires further enhancement, as it lags  in the low-frequency range by 10 orders of magnitude, this indicates the increasing challenge with rising frequency. Our research highlights the potential of spin quantum sensors to investigate fundamental physics at high frequencies.

The investigation into the nature of fundamental constants and their potential variations across space and time constitutes a captivating and multifaceted domain within modern physics, as evidenced by a body of literature~\cite{Uzan:2002vq,Safronova:2017xyt}. 
The prospect of variability in fundamental constants is grounded in factors such as the evolution of background fields or the presence of extra dimensions, concepts often proposed within the realm of particle physics beyond the standard model and string theory. Among the various parameters governing particle physics and gravitational interactions, special attention has been directed toward the fine-structure constant ($\alpha$) and the electron mass ($m_e$) in this sphere of investigation. Precise searches conducted on Earth consistently refine our comprehension of potential variations, narrowing down the permissible limits for such deviations~\cite{RosenbandS}.

The  time-varying fundamental constants manifest in the presence of a background scalar field ~\cite{Graham:2015cka}. This scalar field interacts with the standard model field through the coupling with the electromagnetic field tensor $F_{\mu \nu}$, given by $\propto \phi F_{\mu \nu} F^{\mu \nu}$, and the coupling with the standard model electron field $\psi_e$, given by $\propto \phi m_e \bar{\psi}_e \psi_e$. These interactions lead to  variations in the fine-structure constant $\alpha$ and the electron mass $m_e$, as
\begin{equation}
\alpha(\mathbf{r}, t) = \alpha_0 \left(1 + \frac{1}{\Lambda_{\gamma}} \phi(\mathbf{r}, t)\right),
\end{equation}
\begin{equation}
m_e(\mathbf{r}, t) = m_{e, 0} \left(1 + \frac{1}{\Lambda_{e}} \phi(\mathbf{r}, t)\right).
\end{equation}
Here, $\alpha_0$ and $m_{e, 0}$ denote the bare terms, while $\Lambda_{\gamma}$ and $\Lambda_{e}$ parameterize the respective couplings. 
  
These parameters, $\alpha$ and $m_e$, possess the potential to evolve over cosmological timescales. Ultralight bosons serve as natural dark matter candidates~\cite{2168507}. With   sub-electronvolt (sub-eV) masses, they exhibit substantial occupation numbers, resembling coherent oscillating waves characterized by the frequency $\omega_{\phi}$, leading to scalar field $\phi \propto \cos(\omega_\phi t)$. 
Enhanced field values may emerge when these bosonic waves aggregate into localized compact structures or are emitted from transient astrophysical events~\cite{Dailey:2020sxa}. 
Additionally, topological defects with spatially varying bosonic field values can produce detectable signals when passing through Earth-based detectors~\cite{Afach:2021pfd}.

Temporal variations are characterized prominently by frequency. Signals originating from bosonic fields can cover a broad frequency spectrum, ranging from nHz to PHz for wave-like dark matter. Astrophysical observations, including cosmological assessments and investigations into deviations in strong gravitational potentials, predominantly focus on linear and constant-order variations, probing the lower end of the frequency ranges. 
Existing terrestrial searches for variations in fundamental constants  by utilizing techniques such as atomic spectroscopy~\cite{Zhang:2022ewz} and optomechanical systems~\cite{Vermeulen:2021epa}, have predominantly explored frequencies below $0.1$ GHz.
However, there is currently a dearth of search techniques available for higher frequency ranges.

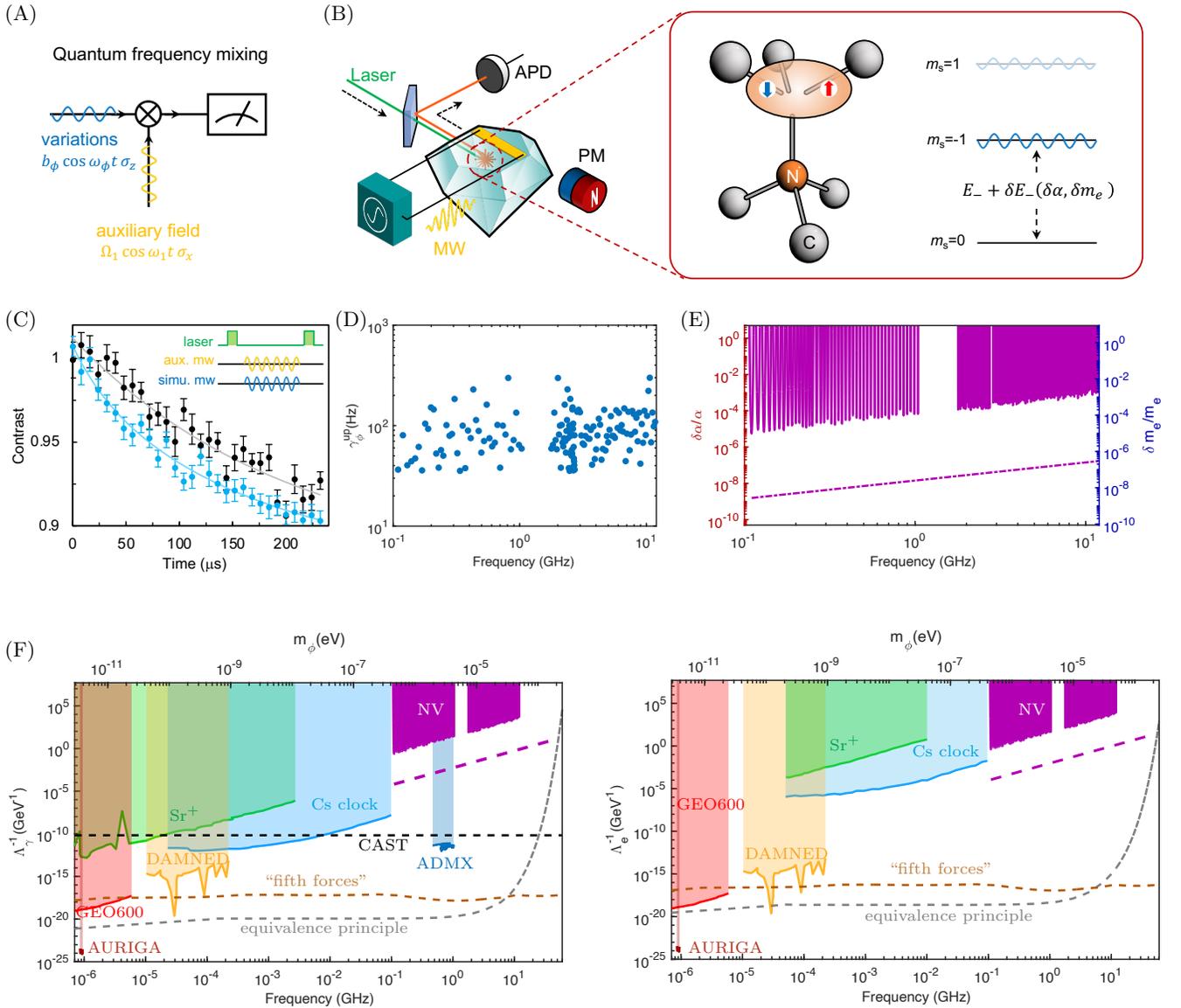
\begin{figure*}[hbtp]
\begin{overpic}[width=1\textwidth]{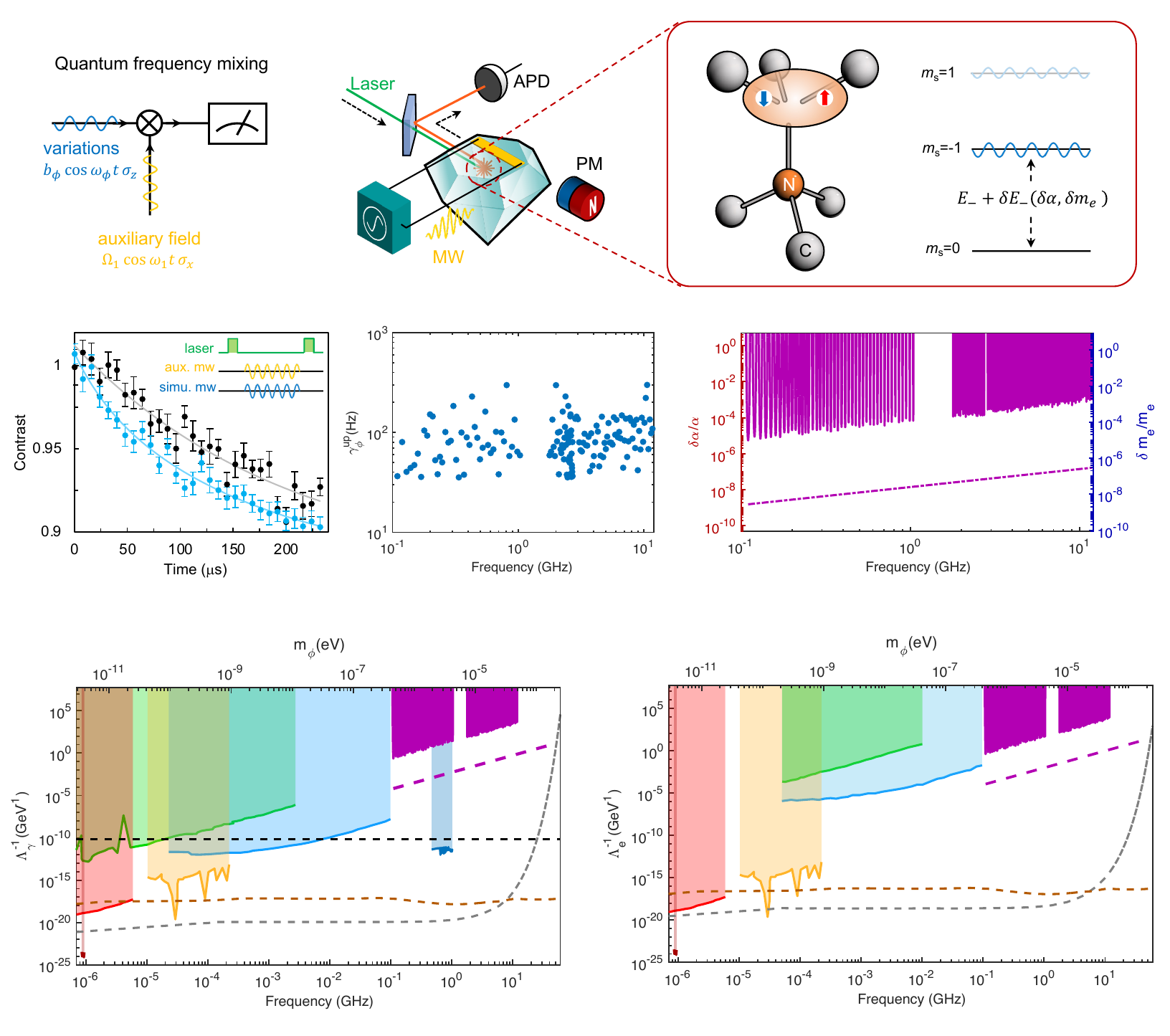} 
  \put (0, 86) { (A)}
  \put (27, 86) { (B)}
  \put (0, 60) { (C)}
  \put (28, 60) { (D)}
  \put (57.3, 60) { (E)}
  \put (0, 32) { (F)}
  \put (6, 9.8) { \textcolor{red}{\scriptsize GEO600}} 
  \put (57, 19) { \textcolor{red}{\scriptsize GEO600}}
  \put (7, 6.8) { \textcolor{BrickRed}{\scriptsize AURIGA}} 
  \put (58, 6.8) { \textcolor{BrickRed}{\scriptsize AURIGA}} 
  \put (12, 14.3) { \textcolor{YellowOrange}{\scriptsize DAMNED}}
  \put (62.8, 14.8) { \textcolor{YellowOrange}{\scriptsize DAMNED}}
  \put (14, 18) { \textcolor{ForestGreen}{\scriptsize Sr$^+$}} 
  \put (70, 24) { \textcolor{ForestGreen}{\scriptsize Sr$^+$}} 
  \put (26, 19) { \textcolor{Cerulean}{\scriptsize Cs clock}} 
  \put (77, 23.5) { \textcolor{Cerulean}{\scriptsize Cs clock}}
  \put (34.9, 14) { \textcolor{NavyBlue}{\scriptsize ADMX}} 
  \put (30, 15.5) { \textcolor{black}{\scriptsize CAST}} 
  \put (20, 8.5) { \textcolor{gray}{\scriptsize equivalence principle}} 
  \put (73, 9.5) { \textcolor{gray}{\scriptsize equivalence principle}}
  \put (22, 12.5) { \textcolor{brown}{\scriptsize ``fifth forces''}}
  \put (75, 13.5) { \textcolor{brown}{\scriptsize ``fifth forces''}} 
  \put (35, 27) { \textcolor{white}{\scriptsize NV}} 
  \put (86, 27) { \textcolor{white}{\scriptsize NV}} 
\end{overpic}
\caption[NV Center]{ (A) The schematic diagram of the experiment. The NV center serves as a quantum mixer to mix the variation signal (blue) with the bias a.c. field (orange) \cite{wang2022Sensing}. (B)
Left: Diamond-based spin quantum sensor is used in our experiment. A green laser initializes the  sensor, and the microwave radiation is fed through a coplanar waveguide for NV spin control. Fluorescence is measured by a confocal optical system. An external permanent magnet adjusts the Zeeman energy level. See details in Supplementary Method II.A.
Middle: Structure of the spin quantum sensor. The NV center is formed by nearest-neighbor pair of a lattice vacancy and a nitrogen atom, which substitutes for a carbon atom. Right: NV center spin energy levels.  The ground state of the NV center is an $S$=1 spin system,  with a zero-field splitting of $D/2\pi=$2.87~GHz. The degenerate $m_s=\pm 1$ states are lift up by a magnetic field, with energy $E_{\pm}/\hbar=D\pm\gamma_eB_0$. Variations in the fundamental constants will lead to changes in the spin levels. (C) We confirm the validity of the quantum mixing method by introducing a simulated signal (see details in Supplementary Method II.B). The black curve represents the spin longitude relaxation without the simulated signal,  whereas the blue curve exhibits the spin longitude relaxation with the simulated signal included, revealing an accelerated decay resulting from quantum mixing process. (D) The upper limit of the additional relaxation rate was determined with 95\% confidence level within the frequency range of 0.1 to 12 GHz. (E) The upper limits of the variations in fundamental constants. Utilizing the upper limits $\gamma_\phi^\text{up}$ of the additional relaxation rate in the frequency range of 0.1-12 GHz, we estimated the upper limit of the variations in fundamental constants, employing a 95\% confidence level. The upper limits of variations in the fine structure constant $\alpha$ at different frequencies are depicted on the left y-axis, while those of variations in the electron mass $m_e$ are displayed on the right y-axis. The estimated results are represented by dashed-dotted lines. The data from our experimental results can be found in Supplementary Table 1. (F) Constraints on the coupling parameters $\Lambda_{\gamma}^{-1}$ and $\Lambda_{e}^{-1}$ as functions of the field mass  $m_{\phi}$ and Compton frequency for scalar field dark matter were established at a 95\% confidence level.  The violet region illustrates the excluded parameter space in this experiment, with the dashed violet line corresponding to simulated results from the ensemble spin sensor. Other colored regions represent excluded parameter spaces  in previous experiments: GEO600  (red), AURIGA  (dark red), DAMNED (orange), dynamical decoupling in Sr$^{+}$ optical clock  (green), Cs clock in cavity  (blue), and ADMX  (navy), CAST search for axions (black dashed line), tests of the equivalence principle (grey dashed line) and searches for ``fifth forces'' (brown dashed line). The data from our experimental results can be found in Supplementary Table 2, while references for other related data are provided in Supplementary Figure S 6.
 }   \label{DM_NV}
\end{figure*}

This study introduces an experiment (Fig.~\ref{DM_NV}(A)) employing a spin quantum sensor to  investigate potential variations in the fine-structure constant $\alpha$ and electron mass $m_e$  using the quantum mixing method \cite{wang2022Sensing}. The spin quantum sensor utilized is the single nitrogen-vacancy (NV) electron spin in diamond (the detailed characteristics of the quantum sensor can be found in Supplementary Method II.C), comprising two strongly coupled electron spins with an effective distance proportional to the cubic lattice spacing $a_0$, which is related to $\alpha$ and $m_e$. 

The NV center, composed of a substitutional nitrogen atom adjacent to a vacancy (Fig.~\ref{DM_NV}(B) left),  
can be described as two holes occupying double-generated orbitals, forming a spin-1 system. 
The energy difference in this system is predominantly determined by the spin-spin interaction $D=2\pi \cdot 2.87$ GHz between these two holes and the Zeeman splitting of spins under an external magnetic field $B_0$. The zero-field splitting $D\propto \mu_B^2a_0^{-3}$ is highly sensitive to variations  in the spin magnetic moment coupling ($\mu_B^2$) and the lattice constant ($a_0$). The former, $\mu_B^2$, is proportional to $\alpha/m_e^2$, while the latter, $a_0$, is proportional to the Bohr radius ($a_B=\hbar /\left(\alpha m_e c\right)$).
Additionally, according to reference \cite{Bloch:2023uis}, the external magnetic field $B_0$ depends on the electron mass, impacting the spin dynamics of the NV center. 
Consequently, the energy difference between the levels of NV center, denoted as $E_{\pm}/\hbar=D\pm\gamma_eB_0$ (Fig.~\ref{DM_NV}(B) right), with $\gamma_e$ the gyromagnetic ratio of electron, provides a valuable means for  searching variations simultaneously in both the electron mass and the fine structure constant. 

In this experiment, we utilize the energy level difference $E_-$ (Fig.~\ref{DM_NV}(B)) to quantify variations in both  the fine structure constant $\alpha$ and the electron mass $m_e$ with $B_0\approx 51$ mT
as (see details in Supplementary Method I.B),
\begin{equation}
\frac{\delta E_{-}}{E_{-}}  \approx \frac{7\delta \alpha}{\alpha}+ \frac{4\delta m_e}{m_e}. \label{E_variance}
\end{equation} 
Through the quantum mixing method \cite{wang2022Sensing}, high-frequency energy level variations can be converted into transitions between NV spin levels (refer to Supplementary Method I.C and I.D). The validity of the method is confirmed by introducing a simulated signal in Supplementary Method II.B. This transition leads to spin relaxation, as depicted in the Fig.~\ref{DM_NV}(C). By measuring the magnitude of this relaxation rate, we can estimate the upper limits on amplitude of the high-frequency variations (see Supplementary Method~II.D for details). Through adjusting the frequency of the bias a.c. field, we can estimate the upper limits of potential variations in fundamental constants (see Supplementary Method~II.E for details) ranging from 0.1 GHz to 12 GHz. The upper limits $\gamma^{\text{up}}_\phi$ of the relaxation rate $\gamma_\phi$ are plotted in Fig.~\ref{DM_NV}(D) at a 95\% confidence level (equivalent to a 2$\sigma$ standard deviation).
It is worth mentioning that our experiment covered the frequency  range of $\omega_\phi/2\pi$ from 0.1 GHz to 12 GHz, comprising a total of 147 experiment points for different $\omega_\phi$ frequencies. The bias a.c. field frequency, $\omega_1/2\pi$, varied approximately from 0.42 GHz to 10.58 GHz  according to Eq. S1, with field strength $\Omega_1/2\pi$ about 10 MHz (refer to Supplementary Figure S~1).

The tests of the relaxation rate $\gamma _\phi$ allow us to estimate the upper limits on magnitudes of high-frequency variations in these constants. 
The high-frequency variations are constrained within a 95\% confidence level, equivalent to a 2$\sigma$ standard deviation. The experimental outcomes are depicted in Fig.~\ref{DM_NV}(E), where the jagged features in the curve result from the near-resonant response of spin quantum sensor, scaling as $\sim \exp(\Delta\omega^2/2\Gamma_2^2)$ (details in Supplementary Method~II.B). The results  demonstrate that within the frequency range of 0.1 GHz to 12 GHz, the best constraint on variations in the fundamental constants lies within the range of approximately $10^{-6}$ to $10^{-3}$ in $\alpha$ and $m_e$.

The essential application of the searches for fundamental constant variations lies in the detection of dark matter.   There is growing attention towards ultra-light dark matter (with mass less than 1 eV).  This type of dark matter is of particular interest as it can address some of the peculiarities of small-scale behavior within galaxies while still being consistent with the success of $\Lambda$CDM (Lambda cold dark matter) on larger scales.
Scalar dark matter, a type of ultra-light dark matter, can interact directly with electrons and electromagnetic fields, leading to changes in the electron mass and the fine-structure constant. The mass of the field is given by  $m_\phi=\hbar\omega_\phi/c^2$, where $\omega_\phi$ represents the angular Compton frequency. 
Currently, direct searches constrain the coupling between scalar field and standard model matter  to below 0.4 $\mu$eV, corresponding to a frequency of 100 MHz. However, there is a current lack of searching methods for higher mass ranges.
By utilizing our method to search fundamental constant variations, we established stringent upper limits on the couplings of scalar dark matter fields in the mass range from 0.4 $\mu$eV to 50 $\mu$eV (0.1 GHz to 12 GHz), as illustrated in Fig.~\ref{DM_NV}(F).

We have evaluated the potential impacts of employing an ensemble of NV centers, considering experimentally feasible parameters such as a 1 mm$^3$ diamond with 10 ppm nitrogen impurities and a 10\% NV yield. The simulation reveals that utilizing an ensemble of NV centers leads to an increased sensitivity in the search for coupling parameters  by a factor of $2.8\times 10^4$. The estimated results are depicted as dash-dotted lines in Fig.~\ref{DM_NV}(F). These estimates suggest that the constraints imposed by the fifth force can be exceeded at 57 GHz (0.24 meV). 
While sensitivity decreases  with an increase in $\omega_\phi$, this effect can be counteracted by enhancing the intensity of the bias a.c. field. Improving the efficiency of the microwave waveguide enables the attainment of approximately the GHz level, thus mitigating the impact of the rising $\omega_\phi$ by at least 3 orders. 
Furthermore, through the utilization of diamond fabrication techniques and quantum control methods, the linewidth of the NV center can be greatly reduced, potentially matching that of the dark matter scalar field, thereby enhancing sensitivity.

In conclusion, we propose and implement a spin quantum sensor based on the NV center in diamond to search the variations in fundamental constants. 
This system operates at significantly higher-frequency ranges beyond the scope of the current gravitational wave detectors and atomic spectroscopy techniques. It enables determination of the upper limit of the  variations of the fine structure constant $\alpha$ and electron mass $m_e$ down to 5 ppm and 8 ppm, respectively, and addresses the search for scalar dark matter fields through their coupling with the standard model fields. 
Additionally, with the advancements in electron spin resonance  technology, a broad frequency range from zero-field  up to at least 10 THz  can be essentially covered. Further developments based on similar approaches will enable extending such systems to cover an even broader frequency spectrum range.

\begin{acknowledgements}

We thank X.\ Rong for helpful discussions.
This work was supported by the National Natural Science Foundation of China (grants No.\ 12150011, No.\ T2388102, No.\ 12075115, No.\ 12075116 and No.\ 11890702), the Anhui Initiative in Quantum Information Technologies (grant No.\ AHY050000) and Innovation Program for Quantum Science and Technology (Grant No. 2021ZD0302200).
Y.C. is supported by VILLUM FONDEN (grant No.\ 37766), by the Danish Research Foundation, and under the European Union’s H2020 ERC Advanced Grant “Black holes: gravitational engines of discovery” grant agreement No. Gravitas–101052587, 
and the Munich Institute for Astro-, Particle and BioPhysics (MIAPbP) which is funded by the Deutsche Forschungsgemeinschaft (DFG, German Research Foundation) under Germany´s Excellence Strategy – EXC-2094 – 390783311,
and by FCT (Fundação para a Ciência e Tecnologia I.P, Portugal) under project No. 2022.01324.PTDC.
\end{acknowledgements}

\end{document}